\documentclass[preprint]{sig-alternate-05-2015}

\begin{document}

\setcopyright{rightsretained}

\doi{}

\isbn{}

\conferenceinfo{Neu-IR '16}{SIGIR Workshop on Neural Information Retrieval, July 21, 2016, 
Pisa, Italy}

\title{LSTM-Based Predictions for Proactive Information Retrieval}

\numberofauthors{1} 
\author{
\alignauthor
Petri Luukkonen,\ Markus Koskela,\ Patrik Flor\'{e}en\\
       \affaddr{Helsinki Institute for Information Technology HIIT}\\
       \affaddr{Department of Computer Science, University of Helsinki}\\
       \email{first.last@helsinki.fi}
} 

\maketitle
\begin{abstract}
\sloppy
  We describe a method for proactive information retrieval targeted at 
  retrieving relevant information during a writing task.
  In our method, the current task and the needs of the user are estimated, 
  and the potential next steps are unobtrusively predicted based on the user's
  past actions.
  We focus on the task of writing, in which the user is coalescing
  previously collected information into a text. Our proactive system
  automatically recommends the user relevant background information.
  The proposed system incorporates text input prediction using a long short-term memory (LSTM) network. 
  We present simulations, which show that the system is able to
  reach higher precision values  in an exploratory search setting compared to both a baseline 
  and a comparison system.
\end{abstract}


%
%



\section{Introduction}

Proactive systems~\cite{Tennenhouse2000} anticipate the needs of the
user and predict possible next steps based on the user's preferences
and current context.  The central component of proactive systems
is the inference engine, which analyzes the current context to provide
the highest-ranking suggestions.  
Proactive systems have recently gained popularity, and 
many of the contemporary major
operating systems include proactive components, e.g., Google Now,
Apple Siri, and Microsoft Cortana.

The rationale for using search engines is to find information that helps us in our daily tasks, be they leisure or professional.
An ideal search engine would support searching and identifying useful information that can then be used in solving these tasks \cite{Vakkari01atheory}. 
In proactive information retrieval~\cite{Bhatia2016} the estimation 
of the current task and context are utilized to proactively retrieve and recommend relevant items.  
In particular, this can include associative forms of recall, e.g.,~in a situation where the 
user does not realize having forgotten about a specific resource~\cite{Rhodes1996}.
A proactive retrieval system can be viewed as a digital personal
assistant that knows the user's preferences and aims to provide useful
and relevant information in the current task context.

Long short-term memory (LSTM) networks \cite{Hochreiter97}
have recently shown remarkable performance in a variety of natural language processing tasks, including speech recognition \cite{Graves2013b}, automatic translation \cite{Sutskever2014}, image captioning \cite{Vinyals2015} and information retrieval \cite{Huang2013,Palangi2015,sordoni2015hierarchical}.
LSTM networks have also been used to generate various kinds of sequences, including text~\cite{Graves2013,sutskever2011generating}. In sequence generation, the network is used to process input sequences one at a time and to sample the next item from the output distribution of the network. The sampled item is then fed to the network as the next input. 
This capability to predict future continuations of a sequence (viz.~text in this case) makes LSTMs particularly attractive for proactive information retrieval. 

\begin{figure}[t]
\centering
\includegraphics[width=\columnwidth, trim={0 15cm 20mm 0cm}]{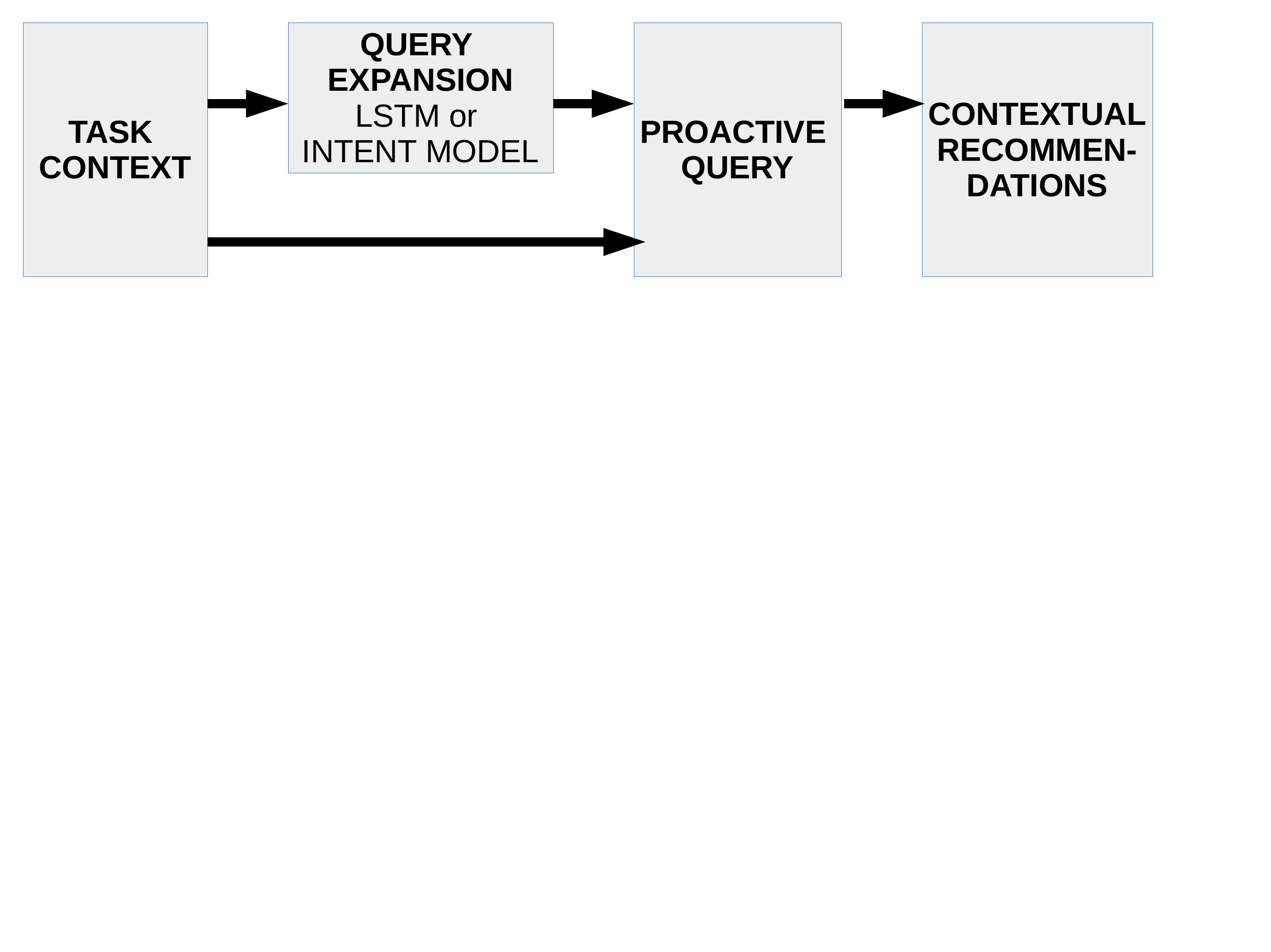}
\caption{Overview of the proposed proactive retrieval system, and the comparison and baseline system used in the experiments.}
\label{fig:system_diagram}
\end{figure}

A typical example of a task is writing about a given topic  \cite{PuertaMelguizo2009,Rhodes1996,Vakkari01atheory}.
In this paper, we propose a method for supporting the user by proactive information  retrieval 
during a writing task. An LSTM network is used to expand the proactive queries by predicting the most likely 
continuations of the current written text. 
An overview of the proposed method is shown in Figure~\ref{fig:system_diagram}.
We present simulated experiments to validate the utility of the LSTM predictions 
in providing relevant documents, and compare the method to a baseline and to another method based on user intent modeling.

The rest of the paper is organized as follows. In the following section a discussion of related work is provided. In Section~\ref{sec:methods} we describe the components 
of our proactive retrieval system: the query expansion methods (based on an LSTM and on user intent modeling), and the proactive query generation.
In Section~\ref{sec:experiment} we describe our simulation experiments.
In Section~\ref{sec:ui} we present our experimental user interface for providing proactive recommendations based on the current task context. The user interface is used in forthcoming 
user studies.
We conclude and discuss future work in Section~\ref{sec:discussion}.


\section{Related Work}\label{sec:rw}
\sloppy

Recurrent neural networks (RNNs), and LSTMs in particular, have recently been used to generate sequences in various domains, such as music~\cite{BoulangerLewandowski2012}, text~\cite{Graves2013,sutskever2011generating}, and handwriting~\cite{Graves2013}.
In information retrieval, RNNs have been used, e.g., for extracting sentence-level semantic vectors~\cite{Palangi2015} and context-aware query suggestion~\cite{sordoni2015hierarchical}.
Other kinds of deep neural networks have been used to project queries and documents to low-dimensional semantic spaces \cite{Huang2013} and to learn fixed-length vectors for variable-length pieces of texts, such as sentences, paragraphs, and documents \cite{Le2014}.

Various types of task activities have been studied in the literature as a basis for query suggestion or query support.
Motivated by the observation that a notable proportion of the user's information needs
were triggered by previous web browsing activity, several authors have studied the correlation between
web browsing behavior and consecutive searches~\cite{cheng2010actively,kong2015predicting,Liebling:2012:ASU:2348283.2348456}.
Another popular approach is to use previous search queries~\cite{Cao:2009:CQC:1571941.1571945,xiang2010context}.
Our work is related to query auto-completion \cite{Bast2006}, in which possible completions of search engine queries are predicted. Query auto-completion for web and mobile queries has recently been studied  in~\cite{mitra2015query,vargas2016term}.

A much studied task scenario for intelligent assistants is writing.
The existing work has largely focused on bibliographic tasks involved in writing scientific or professional texts~\cite{Babaian2002,Livne2014,Twidale2008}. 
These mostly target either the planning or review
phases of writing, as translation, i.e., transforming the writer's mental ideas
into sentences, is the most challenging process in terms of tolerance
to undesired disruptions.
In \cite{PuertaMelguizo2009}, the impact of proactive recommendations
to different phases of the writing process is analyzed.
The \emph{Reactive Keyboard}~\cite{Darragh1990} is an early prototype for predicting the succeeding words from a user-written text fragment.
Completing written sentences has also been studied in \cite{Bickel2005,Grabski2004}.

The setup of our work is closely related to what was envisioned already by Rhodes and Starner in their \emph{Remembrance Agent} (RA)~\cite{Rhodes1996}. The user first indexes her personal data, e.g., emails and written notes. RA is then set to run continuously in the background and display a list of summaries of documents that are related to the current document being read or written.
Other similar tools include \emph{Watson} \cite{Budzik2001} and \emph{Implicit Query} \cite{dumais2004implicit}.
The main difference of our method compared to these systems is the of use an LSTM-based predictive model to perform query expansion.  We also compare our method to a baseline and a method based on user intent modeling.
In this paper, we evaluate our method using a simulated writing task, but our method can be applied to other sources of context, such as what is read on the screen.


\section{Method for Proactive IR}\label{sec:methods}

The proactive recommendations produced by our method are based on user input observed during the current task. 
To improve the recommendations, we propose to use an LSTM-based method for query expansion described in Section~\ref{ssec:lstm}.  For comparison, we use a method based on estimating user intent using a multi-armed bandit model (Section~\ref{ssec:linrel} and Appendix~\ref{sec:linrelappendix}).  Proactive information retrieval using the expanded query is described in Section~\ref{ssec:queryconstr}.
An overview of the proposed method (and the comparison and baseline methods) is shown in Figure~\ref{fig:system_diagram}.

\subsection{LSTM Text Prediction}\label{ssec:lstm}

In LSTM-based query expansion, the most probable  
continuations are estimated for the current written text and 
the expansion words are selected from these estimated 
continuations.
For computing the continuations, we use a beam search algorithm~\cite{Koehn2009} described below.

The LSTM network $f$ is first trained using a text corpus, like abstracts of scientific 
articles. The trained network can then serve as a  
model for text generation.

\begin{figure}
\centering
\includegraphics[width=0.95\columnwidth]{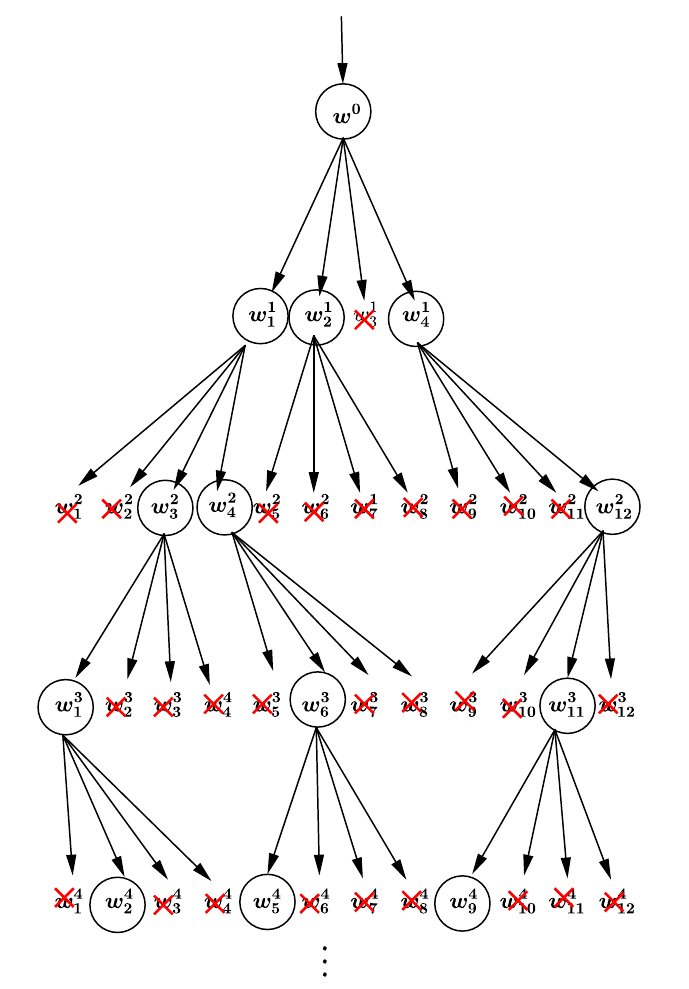}
\caption{A tree is formed from the different continuations estimated from the input word $w^0$. On each level of the tree, nodes other than 
$k=3$ highest scoring ones are pruned out.
As the branching coefficient is $b=4$, there are at most 12
candidates on each level before the pruning.
}
\label{fig:path_formation}
\end{figure}

\begin{figure*}[t]
\centering
\includegraphics[width=0.9\textwidth,trim={0 5mm 0 5mm}]{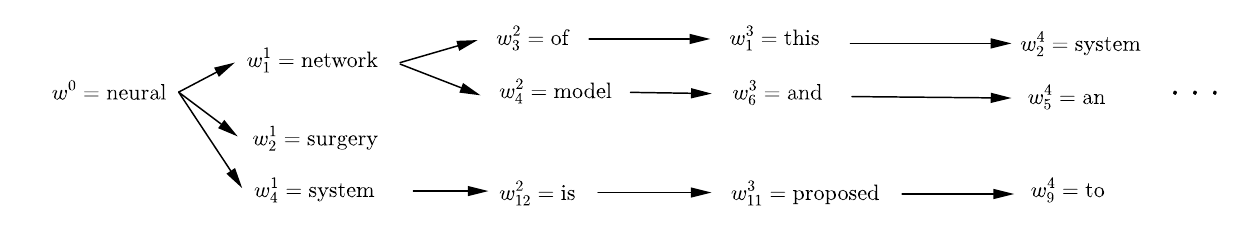}
\caption{Examples of different continuations corresponding the paths in the pruned tree of Figure \ref{fig:path_formation}. 
Words having the highest scores are selected and added to the query. } 
\label{fig:continuations}
\end{figure*}

Let us assume the user has written $n$ input words and denote by $w^0$ the latest word 
in the input sequence.
When given an input word $w$, the network $f$ will 
return a probability distribution for the next word, of which we will consider
the $b$ most probable candidates. 
For example, with the branching coefficient $b=4$ and $w^0 = \text{``neural"}$, the top candidates from $f$ could 
be $\lbrace w_1^1 = 
\text{``model"}, w_2^1 = \text{``network"},  w_3^1 = 
\text{``surgery"}, w_4^1 = \text{``system"} \rbrace $, where $w_i^j$ denotes 
the $i$th candidate for the $j$th word following the input word $w^0$. 
The output probability of the word is denoted as $p(w_i^j)$.
Further continuations are computed by using previous candidate 
words as input to $f$. 
Note that the output of $f$ depends not only on the latest input word, but on all input words so far. 
The different continuations form a tree structure,
where the root node is the input word $w^0$  from 
which the continuations or paths evolve (see Figure~\ref{fig:path_formation}).
We denote by $d$ the depth of the tree, i.e., the number of words in the 
estimated continuations of the sentence.

As the size of the tree grows rather quickly,
the number of generated paths is controlled by the beam width $k$, pruning nodes on each level based on word probabilities on their corresponding paths. 
The path from $w^0$ to $w_i^j$ is denoted by  $\pi(w_i^j)$, e.g., in Figure~\ref{fig:path_formation}, $\pi(w_1^3) = (w^0, w_1^1, w_3^2, w_1^3 )$. The pruning score $R(w_i^j)$ is defined as the product of the word probabilities on $\pi(w_i^j)$: 
\begin{equation}
    R(w_i^j) = \prod_{w' \in \pi(w_i^j)} p(w')\ . 
\end{equation}
On each level of the tree, nodes other than the $k$ highest scoring ones are pruned out.
After the pruning on level $j$, the level $j+1$ candidates are
obtained using the $k$ remaining words. 
Figures \ref{fig:path_formation} and \ref{fig:continuations} show examples of the pruned tree and 
the corresponding continuations.

Continuing our previous example, the second set of estimated words following
the input word $w^0$ are computed using the estimated words 
from the level $1$, e.g.~when feeding to network $f$ the word $w_4^1 = \text{``system"}$, the top candidates returned could be  
$\lbrace w_9^2 = \text{``and"}, w_{10}^2 = \text{``that"}, w_{11}^2
= \text{``of"}, w_{12}^2
= \text{``is"} \rbrace $. 
The third level of generated words are 
computed using the words from the second level and so forth:
$w_{12}^2 = \text{``is"} \rightarrow 
\lbrace w_9^3 = \text{``also"}, w_{10}^3 = \text{``without"},  w_{11}^3
= \text{``proposed"}, w_{12}^3
= \text{``derived"} \rbrace $.

The query expansion is formed from the words remaining in the pruned tree. 
First, the words are filtered using a standard list of English stop words.
The query expansion score of $w_j^i$ is defined 
as the product of its \textit{idf} value and its probability defined
by the model $f$: 
\begin{equation}
\text{score}(w_i^j) = \textit{idf}(w_i^j) \cdot p(w_i^j)\ .
\end{equation}
The \textit{idf} values were computed from the same training data that was used to train the network 
using the form $\textit{idf}(w) = \frac{N}{N_w}$, where 
$N_w$ is the number of documents where the word $w$ appears in, and $N$ is the total number of documents.
Finally, the $n_\text{exp}$ words having the highest scores are added to the expanded query. 

\newpage
\subsection{User Intent Model}\label{ssec:linrel}

As a comparison method, we use an upper confidence bound algorithm, which is based on estimating the user intent using a multi-armed bandit model \cite{glowacka2013directing}. 
The model balances exploration with exploitation and selects words that have the highest upper confidence bound \cite{Auer2003}. This allows the user to interact with words that are relevant, but that are also uncertain to the model. 
Further details of the method are described in Appendix~\ref{sec:linrelappendix}.

\subsection{Proactive Query}
\label{ssec:queryconstr}

Whenever resources are retrieved proactively, our system works as follows; 
for simplicity we refer to this as a \emph{proactive query}, despite that there is no explicit query from the user.
Based on the $n$ input words, our query expansion modules retrieve $n_\text{exp}$ suggested words and 
add them to the query.

The actual information retrieval is performed using the Lucene search engine with the standard cosine-similarity ranking.


\begin{figure*}[t!]
 \centering%
 \includegraphics[width=0.5\textwidth]{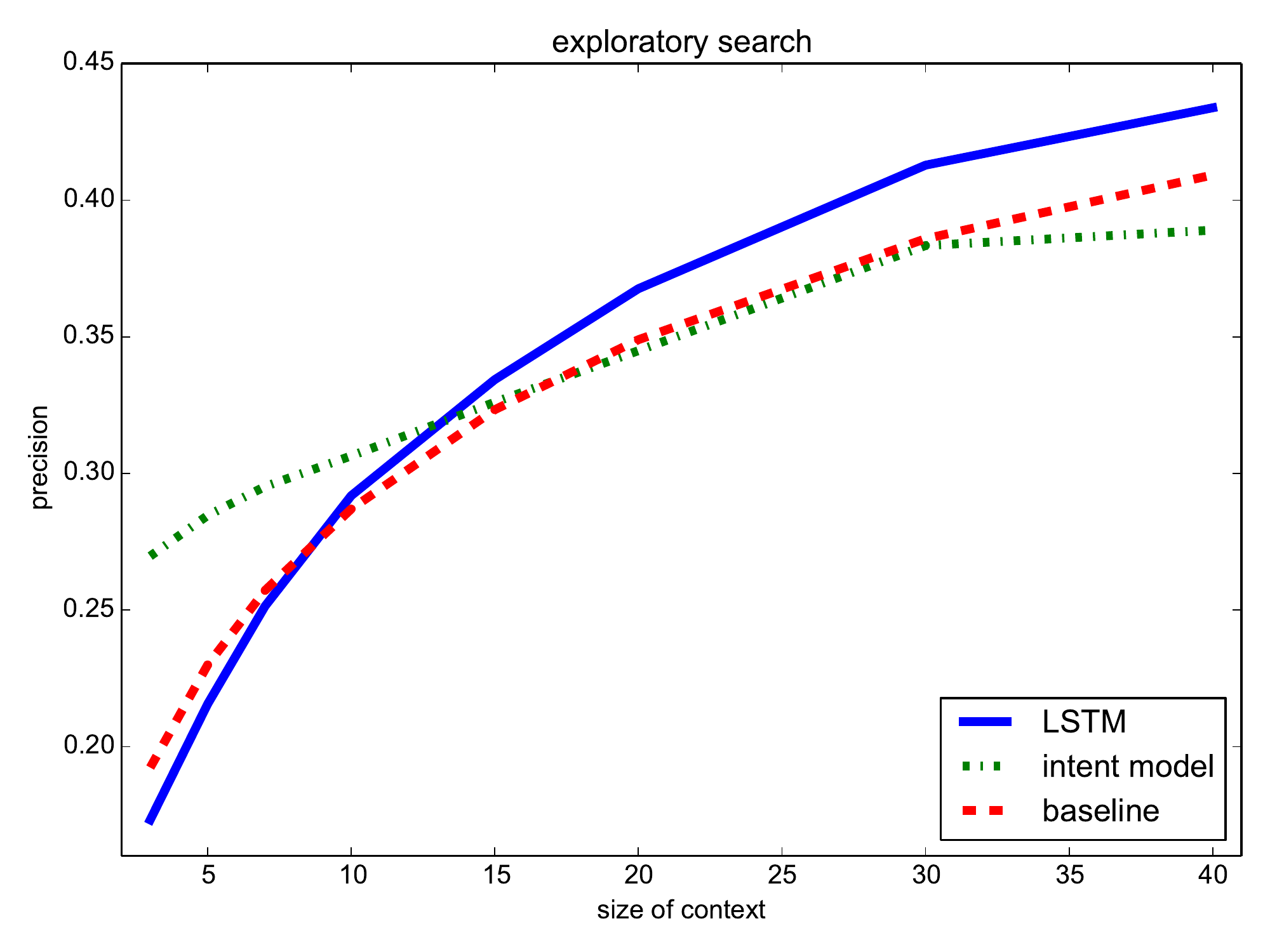}%
 \includegraphics[width=0.5\textwidth]{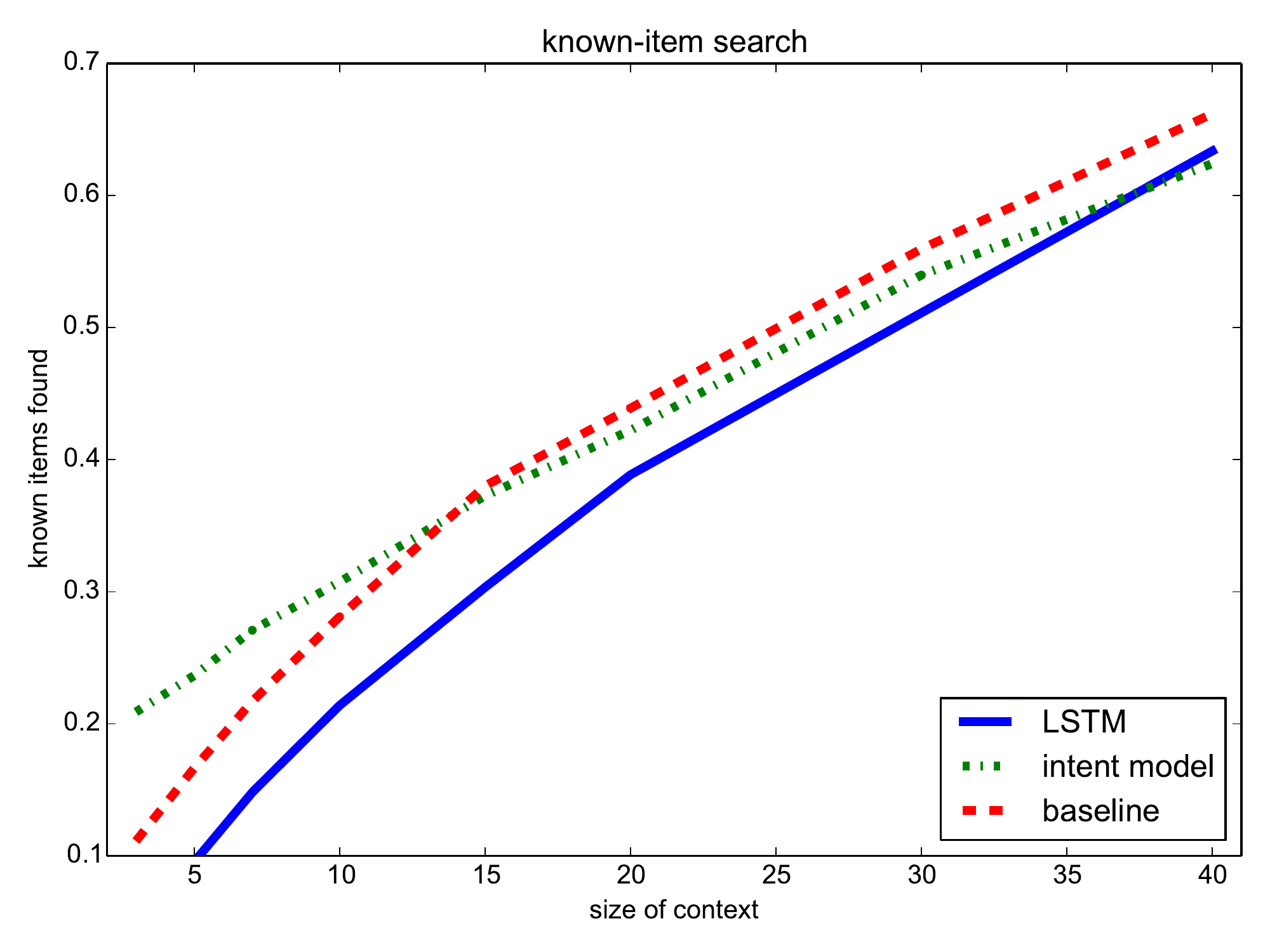}
 
 \caption{Exploratory search document precision (left) and the fraction of known items found (right) for the LSTM, intent model, and baseline runs.
 }
 \label{fig:exp-clicks}
\end{figure*}
 
\section{Simulation Experiments}\label{sec:experiment}

In order to test whether our method provides relevant documents, we performed two kinds of simulation experiments. In the simulations, the input corresponding to text the user types comes from a given document.
We perform query expansion with the LSTM-based text prediction method and with the user intent model.
In the baseline experiments, we use only the written input as the queries.

\subsection{Data Set}
\label{sssec:datasets}

The simulations were performed using the abstracts of the Computer Science branch of the \emph{arXiv}%
\footnote{\url{http://arxiv.org/}}
preprint database, downloaded on October 28, 2015. 
The branch contains a total of 40 subcategories, which are used as topics in our experiments.
A document in the arXiv database can belong to several subcategories, i.e., several topics in our case.

\subsection{Parameters}

We used the \emph{medium} LSTM architecture of Zaremba et al. \cite{Zaremba2014}
with an Tensorflow implementation.
The network has two layers, 650 units per layer, and is unrolled for 35 words.
All the abstracts in the data set were used to train the LSTM network. 
The training data consisted of 15M words with a vocabulary of 10k words.
Training the network took about 36 hours using two GeForce GTX TITAN GPUs.
Due to memory requirements, a random 10\% of the abstracts was used to form the document 
term matrix for the user intent model.

In the LSTM-based word prediction, the beam width 
was set to $k=80$ and the depth to $d=3$.
The branching coefficient was set to $b=10$ and $n_\text{exp}=10$. 

\subsection{Exploratory Search Task}
\label{sssec:sim}

We simulate a setting where a user is writing a text about a given topic.  We choose at random a document and simulate inputting sequences of $n$ consecutive words from the text. The variable $n$ serves as the size of the context expressed as the number of words from the written text. For instance, if $n = 3$, and the test text is 
``Machine learning is a subfield", the input sequences for the proactive search engine would 
be ``Machine learning is", ``learning is a", and ``is a subfield".
We envision a situation where the user has to find some relevant background resources about the given topic. The aim is thus to find other documents about the same topic $t$ as the input document. 
For the proactive queries
(Section~\ref{ssec:queryconstr}), we use $n$ input words and $n_\text{exp}{}=10$.
The Lucene search engine was set to return 10 documents.

We use all the documents in the database 
belonging to the same topic as the test document as the target set $D_t$.
As a measure of performance, we use the \emph{precision} of relevant documents
with regard to the topic $t$ of the input document. 

\subsection{Known-item Search Task}
\label{sssec:sim2}

We also run simulations of known-item search, in which the purpose is to study a setting where the user needs to
re-find a certain previously seen document. 
The setting is the same as for the exploratory search task, except that now we have only one target document in $D_t$.
We take a random document as the query and perform a Lucene search over the rest of the data set to retrieve the highest-scoring document. This is now our target document. Note that the target document is thus a different document than our input document, and in the simulation we either find the target document or not.

\subsection{Results}

Figure~\ref{fig:exp-clicks} shows the results of the simulations. 
On the left-hand side, the retrieval precision in the exploratory search task is shown. The right-hand figure shows the fractions of known items found in the known-item search task. 
First of all, as expected, in both tasks the results improve as the size of the context increases. This was especially anticipated for known-item search, due to how the target document was selected.

Second, the results show that the query expansion methods can improve the precision of the proactively retrieved documents on the exploratory search task. The LSTM-based query expansion improves the results when the context is long enough, i.e., when $n>10$. The intent model based query expansion, on the other hand, is suited for small context sizes ($n<10$).
For known-item search, the query expansion methods are not equally beneficial. The intent model again improves the results for small context sizes, but the LSTM-based predictions degrade the results. 

The simulation results agree rather well with intuition of the query expansion methods. The user intent model
based method expands the query with terms that have high \emph{tf-idf} values in the same documents as the input words. It is conceivable that this is primarily useful when the input context is small, as the expansion can then bring useful additional information to the query.  The LSTM-based query expansion, on the other hand, dynamically models the written context and can predict upcoming words. It is unsurprising that this works better when there is enough input context.  For exploratory search, the predictions made by the LSTM network are accurate enough to increase the retrieval precision; in known-item search the target is smaller, i.e., a single document, and the predictions are not equally useful. 


\begin{figure*}[t!]
\centering
\fbox{\includegraphics[width=0.9\textwidth]{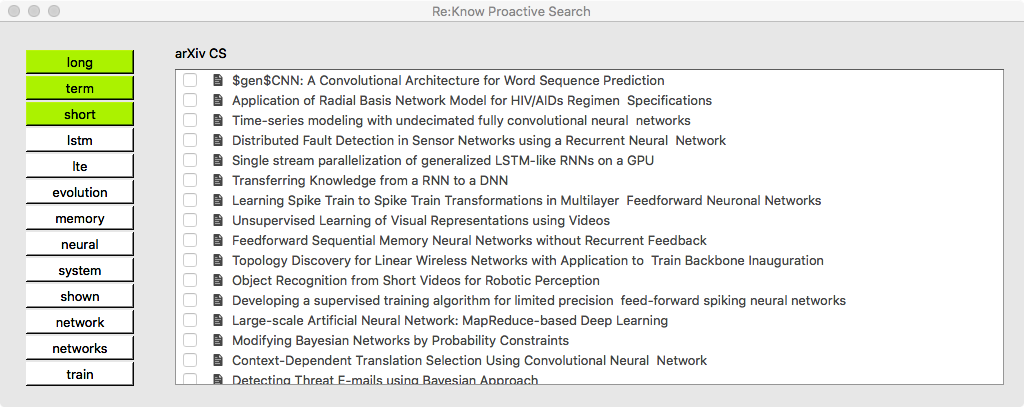}}
\caption{The user interface of the experimental application displaying proactive recommendations. 
The words in the green boxes on the left indicate the written input words (``long short term"). The predictions by the LSTM network are shown in the white boxes.} 
\label{fig:ui}
\end{figure*}

\section{User Interface}
\label{sec:ui}

Our research calls for a user study in order to 
assess the usefulness of the proactive search results in real-world tasks.
For this reason we have implemented an experimental user  
interface intended to be shown in a corner of the screen, 
displaying the 
proactive recommendations; see Figure~\ref{fig:ui}. 
The interface is designed to allow the user to maintain her focus on the current task at hand, while offering peripherally-shown contextual recommendations.
Any data source, e.g., the user's own emails, a database of documents, or any web pages, can be used as information to be proactively retrieved.
In Figure~\ref{fig:ui}, the resources displayed are \emph{arXiv} preprints. 

For obtaining the current writing context, we have implemented a specific
text editor, which transmits the current word surrounding the text cursor at each keypress to the proactive search application; this way the input of $n$ words gradually builds up.
Similarly, the context could be deduced from, e.g., the text read in a web browser.

By clicking on any of the resources, its contents 
(i.e., the corresponding arXiv page in the setting of Figure~\ref{fig:ui}) are shown in a regular web browser.

\section{Discussion and Conclusions}\label{sec:discussion}

We have described a method for query expansion to 
enhance proactive information retrieval.  The method is based on predicting 
the most likely continuations of the current input using an LSTM network.

The performed experiments provide evidence 
that our method is able to proactively produce relevant resources.
The query expansion computed using an LSTM network improved retrieval 
precision in an exploratory search task, when enough context data is available.
The results with the method used as comparison, based on user intent modeling 
using an upper confidence bound algorithm, were partially complementary, improving the 
results when only limited context is available.
This naturally suggests a further study on combining the two query expansion methods.
Further experiments with different kinds of datasets and tasks are in any case needed to validate 
the results.

In this work, we concentrated on the writing task.
We introduced a simple user interface for showing the proactive search results based
on the written context received from a dedicated 
text editor. 
The text predictions produced by the LSTM network could also be used to
to automatically suggest different continuations for the currently written text, 
as in the \emph{Reactive Keyboard}~\cite{Darragh1990}, 
\cite{Bickel2005}, or \cite{Grabski2004}.
Furthermore, user studies where the users are performing real-world tasks need to be carried out.

Finally, in contrast to many of the existing methods for  producing proactive recommendations, the proposed method generalizes the context gathering in the sense that the context data can be extracted from several sources, such as a word processing software, PDF reader, or web browser.
The only requirement for the context data is that it has to be in textual form.  

\section{Acknowledgments}

This work has been partly supported by the Finnish Funding Agency for Innovation (project Re:Know) and the Academy of Finland (Finnish Centre of Excellence in Computational Inference Research COIN, 251170).

{\small

\begin{thebibliography}{10}

\bibitem{Ahukorala2015}
K.~Athukorala, A.~Medlar, K.~Ilves, and D.~Glowacka.
\newblock Balancing exploration and exploitation: Empirical parameterization of
  exploratory search systems.
\newblock In {\em Proc.\ CIKM}, pages 1703--1706, 2015.

\bibitem{Auer2003}
P.~Auer.
\newblock Using confidence bounds for exploitation-exploration trade-offs.
\newblock {\em J. Mach. Learn. Res.}, 3:397--422, Mar. 2003.

\bibitem{Babaian2002}
T.~Babaian, B.~J. Grosz, and S.~M. Shieber.
\newblock A writer's collaborative assistant.
\newblock In {\em Proc.\ IUI}, pages 7--14, 2002.

\bibitem{Bast2006}
H.~Bast and I.~Weber.
\newblock Type less, find more: {F}ast autocompletion search with a succinct
  index.
\newblock In {\em Proc.\ SIGIR}, pages 364--371, 2006.

\bibitem{Bhatia2016}
S.~Bhatia, D.~Majumdar, and N.~Aggarwal.
\newblock Proactive information retrieval: Anticipating users' information
  need.
\newblock In {\em Proc.\ ECIR}, pages 874--877. 2016.

\bibitem{Bickel2005}
S.~Bickel, P.~Haider, and T.~Scheffer.
\newblock Learning to complete sentences.
\newblock In {\em Proc.\ ECML}, pages 497--504. 2005.

\bibitem{BoulangerLewandowski2012}
N.~Boulanger-Lewandowski, Y.~Bengio, and P.~Vincent.
\newblock Modeling temporal dependencies in high-dimensional sequences:
  Application to polyphonic music generation and transcription.
\newblock In {\em Proc. ICML}, pages 1159--1166, 2012.

\bibitem{Budzik2001}
J.~Budzik, K.~Hammond, and L.~Birnbaum.
\newblock Information access in context.
\newblock {\em Knowledge-Based Systems}, 14(1--2):37--53, 2001.

\bibitem{Cao:2009:CQC:1571941.1571945}
H.~Cao, D.~H. Hu, D.~Shen, D.~Jiang, J.-T. Sun, E.~Chen, and Q.~Yang.
\newblock Context-aware query classification.
\newblock In {\em Proc.\ SIGIR}, pages 3--10, 2009.

\bibitem{cheng2010actively}
Z.~Cheng, B.~Gao, and T.-Y. Liu.
\newblock Actively predicting diverse search intent from user browsing
  behaviors.
\newblock In {\em Proc. WWW}, pages 221--230, 2010.

\bibitem{Darragh1990}
J.~J. Darragh, I.~H. Witten, and M.~L. James.
\newblock {The Reactive Keyboard: A} predictive typing aid.
\newblock {\em Computer}, 23(11):41--49, 1990.

\bibitem{dumais2004implicit}
S.~Dumais, E.~Cutrell, R.~Sarin, and E.~Horvitz.
\newblock Implicit queries ({IQ}) for contextualized search.
\newblock In {\em Proc.\ SIGIR}, pages 594--594, 2004.

\bibitem{glowacka2013directing}
D.~Glowacka, T.~Ruotsalo, K.~Konuyshkova, S.~Kaski, G.~Jacucci, et~al.
\newblock Directing exploratory search: Reinforcement learning from user
  interactions with keywords.
\newblock In {\em Proc. IUI}, pages 117--128, 2013.

\bibitem{Grabski2004}
K.~Grabski and T.~Scheffer.
\newblock Sentence completion.
\newblock In {\em Proc.\ SIGIR}, pages 433--439, 2004.

\bibitem{Graves2013}
A.~Graves.
\newblock Generating sequences with recurrent neural networks.
\newblock {\em arXiv preprint arXiv:1308.0850}, 2013.

\bibitem{Graves2013b}
A.~Graves, A.~r.~Mohamed, and G.~Hinton.
\newblock Speech recognition with deep recurrent neural networks.
\newblock In {\em Proc ICASSP}, pages 6645--6649, 2013.

\bibitem{Hochreiter97}
S.~Hochreiter and J.~Schmidhuber.
\newblock Long short-term memory.
\newblock {\em Neural Computation}, 9(8):1735--1780, 1997.

\bibitem{Huang2013}
P.-S. Huang, X.~He, J.~Gao, L.~Deng, A.~Acero, and L.~Heck.
\newblock Learning deep structured semantic models for web search using
  clickthrough data.
\newblock In {\em Proc.\ CIKM}, pages 2333--2338, 2013.

\bibitem{Koehn2009}
P.~Koehn.
\newblock {\em Statistical machine translation}.
\newblock Cambridge University Press, 2009.

\bibitem{kong2015predicting}
W.~Kong, R.~Li, J.~Luo, A.~Zhang, Y.~Chang, and J.~Allan.
\newblock Predicting search intent based on pre-search context.
\newblock In {\em Proc.\ SIGIR}, pages 503--512, 2015.

\bibitem{Le2014}
Q.~V. Le and T.~Mikolov.
\newblock Distributed representations of sentences and documents.
\newblock {\em arXiv preprint arXiv:1405.4053}, 2014.

\bibitem{Liebling:2012:ASU:2348283.2348456}
D.~J. Liebling, P.~N. Bennett, and R.~W. White.
\newblock Anticipatory search: Using context to initiate search.
\newblock In {\em Proc.\ SIGIR}, pages 1035--1036, 2012.

\bibitem{Livne2014}
A.~Livne, V.~Gokuladas, J.~Teevan, S.~T. Dumais, and E.~Adar.
\newblock Citesight: Supporting contextual citation recommendation using
  differential search.
\newblock In {\em Proc.\ SIGIR}, pages 807--816, 2014.

\bibitem{PuertaMelguizo2009}
M.~C.~P. Melguizo, L.~Boves, and O.~M. Ramos.
\newblock A proactive recommendation system for writing: Helping without
  disrupting.
\newblock {\em International Journal of Industrial Ergonomics}, 39(3):516--523,
  2009.

\bibitem{mitra2015query}
B.~Mitra and N.~Craswell.
\newblock Query auto-completion for rare prefixes.
\newblock In {\em Proc.\ CIKM}, pages 1755--1758, 2015.

\bibitem{Palangi2015}
H.~Palangi, L.~Deng, Y.~Shen, J.~Gao, X.~He, J.~Chen, X.~Song, and R.~Ward.
\newblock Deep sentence embedding using long short-term memory networks.
\newblock {\em arXiv:1502.06922}, 2015.

\bibitem{Rhodes1996}
B.~Rhodes and T.~Starner.
\newblock {R}emembrance {A}gent: A continuously running automated information
  retrieval system.
\newblock In {\em Proc.\ PAAM96}, pages 487--495, 1996.

\bibitem{sordoni2015hierarchical}
A.~Sordoni, Y.~Bengio, H.~Vahabi, C.~Lioma, J.~Grue~Simonsen, and J.-Y. Nie.
\newblock A hierarchical recurrent encoder-decoder for generative context-aware
  query suggestion.
\newblock In {\em Proc.\ CIKM}, pages 553--562, 2015.

\bibitem{sutskever2011generating}
I.~Sutskever, J.~Martens, and G.~E. Hinton.
\newblock Generating text with recurrent neural networks.
\newblock In {\em Proc.\ ICML}, pages 1017--1024, 2011.

\bibitem{Sutskever2014}
I.~Sutskever, O.~Vinyals, and Q.~V. Le.
\newblock Sequence to sequence learning with neural networks.
\newblock {\em arXiv preprint arXiv:1409.3215}, 2014.

\bibitem{Tennenhouse2000}
D.~Tennenhouse.
\newblock Proactive computing.
\newblock {\em Commun. ACM}, 43(5):43--50, May 2000.

\bibitem{Twidale2008}
M.~B. Twidale, A.~A. Gruzd, and D.~M. Nichols.
\newblock Writing in the library: Exploring tighter integration of digital
  library use with the writing process.
\newblock {\em Information Processing \& Management}, 44(2):558--580, 2008.

\bibitem{Vakkari01atheory}
P.~Vakkari.
\newblock A theory of the task-based information retrieval process: a summary
  and generalization of a longitudinal study.
\newblock {\em Journal of Documentation}, 57(1):44--60, 2001.

\bibitem{vargas2016term}
S.~Vargas, R.~Blanco, and P.~Mika.
\newblock Term-by-term query auto-completion for mobile search.
\newblock In {\em Proc.\ WSDM}, pages 143--152, 2016.

\bibitem{Vinyals2015}
O.~Vinyals, A.~Toshev, S.~Bengio, and D.~Erhan.
\newblock Show and tell: A neural image caption generator.
\newblock In {\em Proc. CVPR}, pages 3156--3164, 2015.

\bibitem{xiang2010context}
B.~Xiang, D.~Jiang, J.~Pei, X.~Sun, E.~Chen, and H.~Li.
\newblock Context-aware ranking in web search.
\newblock In {\em Proc. SIGIR}, pages 451--458, 2010.

\bibitem{Zaremba2014}
W.~Zaremba, I.~Sutskever, and O.~Vinyals.
\newblock Recurrent neural network regularization.
\newblock {\em arXiv preprint arXiv:1409.2329}, 2014.

\end{thebibliography}

}

\appendix
\section{User Intent Model}
\label{sec:linrelappendix}

For computing the user intent model, we use a training database consisting of $M$ 
documents, from which $N$ unique words are extracted by excluding stop words. 
The $j$th document in the database is represented by a feature vector 
$x_j \in \mathbb{R}^{N}$ where
$x_{ij}$ is the \textit{tf-idf} value of the $i$th word.
We denote by $X \in \mathbb{R}^{N \times M}$ the \textit{tf-idf}
matrix of the $M$ documents, where each column of $X$
corresponds to one document feature vector and each row corresponds to a distribution of the words over the documents.

The user intent model is estimated using
the context formed using $n$ previously written words.
Based on this input, a set of word weights are computed by using the LinRel algorithm 
proposed in~\cite{Auer2003}.

We denote the relevance vector of observed words by 
$y \in [0, 1]^{N}$ , 
where $y_i=1$ corresponds to having observed the
$i$th word in the input. 
If the $i$th word does not occur in the subsequent input words, its relevance value 
starts decreasing such that $y_i = n_i^{-1}$, where $n_i$ is the 
number of sets of input words since the last occurrence of the $i$th word. 
For omitting words having very low relevance values, we 
use a threshold $\tau =0.1$: when $y_i < \tau$
the value of $y_i$ is set to zero.

The observed values in $y$, corresponding to the input words so far,
are assumed to be formed from the model 
\begin{equation}
y  = X \hat{w}\ ,
\end{equation}
where $\hat{w} \in \mathbb{R}^{M}$ is the estimated \emph{user intent model} 
describing what documents from the training set are currently
estimated to be relevant for the user.

Given $y$ and $X$, the user model $\hat{w}$ can be obtained as
\begin{equation} \label{eq:tikh}
	\hat{w} = (X^T X + \mu I)^{-1}X^T y \ ,
\end{equation}
where $I$ is an identity matrix of size $M \times M$ and $\mu \geq 0$ is a regularization parameter, set to $\mu = 1.0$ in our experiments.

Using Eq.~(\ref{eq:tikh}) the relevance estimate $\hat{y}$ of the
words in the vocabulary is computed as
\begin{equation}
\hat{y} = X\hat{w} = X (X^T X + \mu I)^{-1}X^T y = Ay
\end{equation}
and the upper bound of the standard deviation of $\hat{y}_i$ as
\begin{equation}
\hat{\sigma}_i = \|row_i(A)\|^2.
\end{equation}

The $n_\text{exp}$ words to be included in the expanded query correspond to the $n_\text{exp}$ maximum
components of the vector 
\begin{equation} \label{eq:v_hat}
v = \hat{y} + c\hat{\sigma},
\end{equation}
with words appearing in the input excluded. Here $c \geq 0$ is the exploration/exploitation parameter
controlling the trade-off between exploring the search 
space (large $c$) and focusing on the currently most promising region (small $c$). 
We use here $c=1.0$ as recommended in the literature~\cite{Ahukorala2015}.

\end{document}